\documentclass[aps,prd,groupaddress,tighten,nofootinbib,superscriptaddress,showpacs]{revtex4}
\usepackage{amssymb}
\usepackage{amsmath,color}
\usepackage{epsfig}
\usepackage{graphicx,epsf,epsfig}
\usepackage{bbm}
\usepackage{footnote}

\usepackage{color}

\def\slc#1{\setbox0=\hbox{$#1$}           
    \dimen0=\wd0                                 
    \setbox1=\hbox{/} \dimen1=\wd1               
    \ifdim\dimen0>\dimen1                        
       \rlap{\hbox to \dimen0{\hfil/\hfil}}      
       #1                                        
    \else                                        
       \rlap{\hbox to \dimen1{\hfil$#1$\hfil}}   
       /                                         
    \fi}

\def\be{\begin{equation}}
\def\ee{\end{equation}}
\def\gs{\mathrel{
   \rlap{\raise 0.511ex \hbox{$>$}}{\lower 0.511ex \hbox{$\sim$}}}}
\def\ls{\mathrel{
   \rlap{\raise 0.511ex \hbox{$<$}}{\lower 0.511ex \hbox{$\sim$}}}}

\newcommand{\onbb}{neutrinoless double beta decay }
\newcommand{\ba}{\begin{array}{c}}
\newcommand{\baz}{\begin{array}{cc}}
\newcommand{\barrr}{\begin{array}{rrr}}
\newcommand{\bad}{\begin{array}{ccc}}
\newcommand{\bav}{\begin{array}{cccc}}
\newcommand{\baf}{\begin{array}{ccccc}}
\newcommand{\bea}{\begin{equation} \begin{array}{c}}
\newcommand{\eea}{ \end{array} \end{equation}}
\newcommand{\ea}{\end{array}}

\newcommand{\meff}{\mbox{$\langle m \rangle$}}

\begin{document}
\title{Symmetrical Parametrizations of the Lepton Mixing Matrix
}
\date{\today}
\author{W.~Rodejohann}
\email{werner.rodejohann@mpi-hd.mpg.de}

\affiliation{Max-Planck-Institut f{\"u}r Kernphysik, Postfach
103980, 69029 Heidelberg, Germany}

\author{J.~W.~F.~Valle}
\email{valle@ific.uv.es}

\affiliation{AHEP Group, Institut de F\'{\i}sica Corpuscular --
C.S.I.C./Universitat de Val{\`e}ncia, 
Edificio Institutos de Paterna, Apt 22085, E--46071 Valencia, Spain}

\pacs{14.60.Lm, 14.60.Pq}

\begin{abstract}
  Advantages of the original symmetrical form of the parametrization
  of the lepton mixing matrix are discussed. It provides a
  conceptually more transparent description of neutrino oscillations
  and lepton number violating processes like neutrinoless double beta
  decay, clarifying the significance of Dirac and Majorana phases. It
  is also ideal for parametrizing scenarios with light sterile
  neutrinos.
\end{abstract}
\maketitle

\section{Introduction}
Since the historic discovery of neutrino oscillations, massive
neutrinos currently provide the most direct and testable evidence for
physics beyond the Standard Model (SM) of particle physics.
At low energies, nine (seven) parameters must
be determined depending on whether neutrinos are Majorana (Dirac)
particles~\cite{schechter:1980gr}\footnote{Many more parameters exist
  if neutrino masses arise from type-I seesaw, the parametrization in
  Ref.~\cite{schechter:1980gr} covers all seesaw cases.}. Here we
tacitly assume the former, more general, and theoretically preferred
case.
Parametrizing the lepton mixing matrix \cite{P,MNS} in a convenient
and intuitive manner is very helpful for data handling and
interpretation of the physics, such as neutrino oscillation searches
in upcoming long-baseline
experiments~\cite{bandyopadhyay:2007kx,Nunokawa:2007qh} or searches
for neutrinoless double beta
decay~\cite{avignone:2007fu,Rodejohann:2011mu}.
While the former are sensitive to the Dirac phase, Majorana
phases~\cite{Bilenky:1980cx,schechter:1980gr,schechter:1980gk,Doi:1980yb,degouvea:2002gf}
are crucial to describe the latter.

The Particle Date Group (PDG) has adopted a parametrization of the
lepton mixing matrix in which it is a product of three consecutive
rotations multiplied with a diagonal phase matrix $P$ containing the
Majorana phases \cite{nakamura2010review}.  The mixing matrix can be
written as
\begin{equation}
 \label{eq:pdg} U
= R_{23}(\theta_{23}; 0) \, R_{13}(\theta_{13}; \delta) \,
R_{12}(\theta_{12}; 0) \, P \, , \ee where $R_{ij}(\theta; \varphi)$
is a rotation around the $ij$-axis, e.g.  \begin{equation} \label{eq:rot}
R_{13}(\theta_{13}; \delta) = \left( \bad
  \cos \theta_{13} & 0 & \sin \theta_{13} \, \text{e}^{-i \delta}   \\
  0 & 1 & 0 \\
  -\sin \theta_{13} \, \text{e}^{i \delta} & 0 & \cos \theta_{13} \ea
\right) .  
  \end{equation}
  The position of the Dirac phase $\delta$ is the convention chosen by the PDG. 
  The two Majorana phases, denoted here $\alpha$ and
  $\beta$, are usually put inside $P$, to the right of the mixing matrix: 
  $P = {\rm diag}(\text{e}^{i\alpha},\text{e}^{i\beta},1)$. However there is no 
  consensus notation yet in what concerns the parametrization of these phases, neither for 
  their names (e.g.~$\phi_1$ and $\phi_2$ or $\varphi_1$ and $\varphi_2$ or 
  $\sigma$ and $\rho$, sometimes with a minus sign, sometimes divided by two, etc.), 
  nor for their position within the matrix ($P = {\rm diag}(1,\text{e}^{i\alpha},\text{e}^{i\beta})$, or $P =
  {\rm diag}(\text{e}^{i\alpha},\text{e}^{i\beta},1)$, etc.).  The
  mixing matrix Eq.~(\ref{eq:pdg}) is explicitly given by
\begin{widetext}
\begin{equation} 
U \equiv   \tilde{U} \, P  = 
\label{eq:pdg_full}  
 \left( \bad 
c_{12}  c_{13} \, \text{e}^{i\alpha} & s_{12}  c_{13} \, \text{e}^{i\beta} 
& s_{13}  \, \text{e}^{-i \delta}  \\ 
-(s_{12}  c_{23} + c_{12}  s_{23}  s_{13}  \, \text{e}^{i \delta})\, \text{e}^{i\alpha} 
& (c_{12}  c_{23} - s_{12}  s_{23} s_{13}  \, \text{e}^{i \delta})  \, \text{e}^{i\beta}
& s_{23}  c_{13}  \\ 
(s_{12}  s_{23} - c_{12}  c_{23}  s_{13}  \, \text{e}^{i \delta}) \, \text{e}^{i\alpha} & 
- (c_{12}  s_{23} + s_{12}  c_{23}  s_{13}  \, \text{e}^{i \delta} ) \, \text{e}^{i\beta}
& c_{23} c_{13}  
\ea   
\right) \,.  
 \end{equation}
 \end{widetext}

 The elements $|U_{e1}|$ and $|U_{\mu 3}|$ are known with good accuracy, thus 
 the two "large" angles $\theta_{12}$ and $\theta_{23}$ are well-determined by 
 solar and atmospheric neutrino oscillation data. There is also recent 
 evidence for a nonzero value of the "small" element $|U_{e3}|$ from the
 T2K Collaboration~\cite{Abe:2011sj} and the global neutrino oscillation data 
 sample~\cite{Schwetz:2011qt,Fogli:2011qn}. 
While there are 8 equivalent ways to parametrize the mixing 
matrix~\cite{Fritzsch:2001ty}, the above order of rotation is useful in 
the sense that experimentally the straightforwardly measurable elements 
$|U_{e1}|$, $|U_{e3}|$ and $|U_{\mu 3}|$ allow to directly extract the angles 
in the three--neutrino lepton mixing matrix~\footnote{See Ref.~\cite{Rodejohann:2011uz} for a recent application 
  of the other possible parametrizations, and \cite{ahn} for a 
rare use of the symmetrical parametrization we will study in this
work.}. In contrast, the other equivalent 
parametrizations do not share this property. 
 
\section{A more convenient parametrization of the mixing 
matrix}

The above form $U$ is nothing but a re-writing of the ``symmetrical'' 
form $K$ proposed in Ref.~\cite{schechter:1980gr} (apart from factor 
ordering, which was left unspecified in the original paper). 
Here we would like to argue in favor of the conceptual advantages of 
the original ``symmetrical''  presentation of the lepton mixing 
matrix~\cite{schechter:1980gr}. 
For the case of three neutrinos it is given as: 
\begin{equation} \label{eq_jv}
K = \omega_{23}(\theta_{23}; \phi_{23}) \omega_{13}(\theta_{13}; \phi_{13})
\omega_{12}(\theta_{12}; \phi_{12}) \, , 
  \end{equation}
where each of the $\omega$'s is effectively $2\times 2$, characterized by 
an angle and a CP phase, e.g.
\begin{equation}
   \label{eq:w13}
\omega_{13} = \left(\begin{array}{ccccc}
c_{13} & 0 & e^{-i \phi_{13}} s_{13} \\
0 & 1 & 0 \\
-e^{i \phi_{13}} s_{13} & 0 & c_{13}
\end{array}\right)\,.  \nonumber
 \end{equation}
Explicitly, the symmetrical parametrization of the lepton mixing matrix, $K$
 can be written as:
\begin{equation}
K=\left( \begin{array}{c c c}
c_{12}c_{13}&s_{12}c_{13}e^{-i{\phi_{12}}}&s_{13}e^{-i{\phi_{13}}}\\
-s_{12}c_{23}e^{i{\phi_{12}}}-c_{12}s_{13}s_{23}e^{-i({\phi_{23}}-{\phi_{13}})}
&c_{12}c_{23}-s_{12}s_{13}s_{23}
e^{-i({\phi_{12}}+{\phi_{23}}-{\phi_{13}})}&c_{13}s_{23}e^{-i{\phi_{23}}}\\
s_{12}s_{23}e^{i({\phi_{12}}+{\phi_{23}})}-c_{12}s_{13}c_{23}e^{i{\phi_{13}}}
&-c_{12}s_{23}e^{i{\phi_{23}}}-
s_{12}s_{13}c_{23}e^{-i({\phi_{12}}-{\phi_{13}})}&c_{13}c_{23}\\
\end{array} \right).
\label{writeout}
\end{equation}
Here all three CP violating phases 
are physical~\cite{schechter:1980gk}: $\phi_{12},\phi_{23}$ and $\phi_{13}$. 
Even though the parametrization is fully ``symmetric'' there is a basic 
difference between Dirac and Majorana phases. In order to understand this
let us use the identity~\cite{schechter:1980gr} 
\begin{equation}
P^{-1} \, K \, P = 
 \omega_{23}(\theta_{23}; \phi_{23} - \beta)~
\omega_{13}(\theta_{13}; \phi_{13} - \alpha)~\omega_{12}(\theta_{12};
\phi_{12} + \beta -\alpha) \,,
  \end{equation}
 which allows us, up to unphysical phases, to identify the Dirac phase as 
\begin{equation} \label{eq:delta}
\delta \leftrightarrow \phi_{13} -\phi_{12} - \phi_{23} \, , 
\end{equation}

This formula relates the Dirac phase, which denotes the phase responsible
for CP violation in neutrino oscillations in the PDG parametrization,
with the equivalent phase in the symmetrical representation. 
Note that it obeys field rephasing invariance, as it should.
Moreover, in contrast to the PDG description, in the symmetrical form 
CP violation in neutrino oscillations is immediately recognized as a 
three-generation phenomenon involving the phases of all three 
generations\footnote{In this sense the Dirac phase has 
  an intrinsic geometric meaning, like the curl of two vectors or the area of 
  a triangle.}, an important conceptual advantage.

Recall that the neutrino oscillation probability for a 
$\nu_\alpha \to \nu_\beta$ flavor transition is given by 
\begin{eqnarray}
P(\nu_\alpha \to  \nu_\beta)   =  
\left| \sum_j U^*_{\alpha j} U_{\beta j} \,  
\text{e}^{-i\frac{m^2_j}{2E}L} \right|^2\nonumber 
 =  \delta_{\alpha \beta} 
-4 \sum_{i>j} {\Re}\left\{ U^*_{\alpha i} U_{\alpha j} U_{\beta i} U^*_{\beta j} \right\}
\sin^2\left(  \frac{\Delta m^2_{ij}}{4E}L \right)  \\
   +2 \sum_{i>j} 
{\Im} \left\{U^*_{\alpha i} U_{\alpha j} U_{\beta i} U^*_{\beta j} \right\}
\sin \left( \frac{\Delta m^2_{ij}}{2E}L \right),\nonumber
\label{eq:prob_vac}
\end{eqnarray}
where $E$ is the neutrino energy, $L$ is the distance traveled by the 
neutrino, and $\Delta m_{ij}^2 \equiv m_i^2 - m_j^2$ ($m_i$ being
positive mass eigenvalues) are the neutrino mass-squared differences. 
Here $\Re$ and $\Im$ denote real and imaginary parts. For three families 
there is only one independent imaginary part 
${\Im} \left\{U^*_{\alpha i} U_{\alpha j} U_{\beta i} U^*_{\beta j}\right\}$, 
which is responsible for CP violation in neutrino oscillations. 
Comparing this invariant with the PDG and symmetrical
parametrizations gives  
\begin{eqnarray} \nonumber 
J_{\rm CP} & = & {\Im}\left\{ U_{e1}^\ast \, U_{\mu 3}^\ast \, U_{e3} \,
U_{\mu 1} \right\} = 
\left\{ 
\baz  
\frac 18  \sin 2 \theta_{12} \, \sin 2 \theta_{23} \, \sin 2
\theta_{13} \, \cos \theta_{13} \, \sin \delta & \mbox{(PDG)} \, ,
\\
\frac 18  \sin 2 \theta_{12} \, \sin 2 \theta_{23} \, \sin 2
\theta_{13} \, \cos \theta_{13} \, \sin (\phi_{13} - \phi_{12}
- \phi_{23}) & \mbox{(symmetrical)} \, ,
\ea 
\right. 
\end{eqnarray}
and shows the same result as Eq.~(\ref{eq:delta}).

\subsection{Application to Lepton Number Violating Phenomena}
As well known, there are two conceptually different kinds of CP 
violating phenomena~\cite{schechter:1980gr}. 
In the language of the PDG parametrization, one is associated to 
the ``Dirac phase'' $\delta$ and is the exact analogue to the CP phase 
in the quark mixing matrix, responsible for the area of the 
Cabibbo-Kobayashi-Maskawa (CKM) unitarity triangle; while the 
other one is associated to the two ``Majorana phases'' 
$\alpha$ and $\beta$, which do not show up in neutrino 
oscillations~\cite{Bilenky:1980cx,schechter:1980gk,Doi:1980yb,degouvea:2002gf}
but do affect lepton number violating amplitudes. 

In what follows we will briefly discuss the role of Majorana phases 
in determining the rates characterizing neutrinoless double beta decay 
and neutrino-anti-neutrino oscillations~\cite{schechter:1980gk}. 

A suitable parametrization of Majorana phases plays a very important role  
in interpreting the effective mass parameter characterizing the amplitude 
for neutrinoless double beta decay (see Ref.~\cite{Rodejohann:2011mu} for 
a recent review). Its explicit form reads 
\begin{equation} \label{eq:meff_expl}
\meff = \left|\sum_j U_{ej}^2 \, m_j \right| 
= 
\left\{ 
\baz \left| c_{12}^2 c_{13}^2  \,m_1 \, 
\text{e}^{2 i \alpha} + s_{12}^2 c_{13}^2 \, m_2 \, \text{e}^{2 i \beta}+
s_{13}^2  \,m_3 \,\text{e}^{2 i \delta} \right| 
& \mbox{(PDG)} \, , \\ 
\left| c_{12}^2 c_{13}^2  \,m_1 + s_{12}^2 c_{13}^2  \,m_2 \,\text{e}^{2 i \phi_{12}}+
s_{13}^2  \, m_3  \,\text{e}^{2 i \phi_{13}}
\right| & \mbox{(symmetrical)} \, . 
 \ea \right.    
  \end{equation}
  Only the two Majorana phases should appear in \meff~\cite{schechter:1980gk}. 
  However this is not at all clear in the PDG presentation. In contrast, 
  the symmetrical parametrization provides a manifestly transparent 
  description in which only the two Majorana phases appear in \meff, 
  as it should. 
  Currently nuclear matrix element uncertainties prevent the extraction of 
  Majorana phases from neutrinoless double beta decay. However, should these be 
  circumvented and should the determination of the Majorana phases become an 
  issue, then the symmetrical parametrization will surely be preferred over 
  the PDG one. 

  It has long been known that the lepton mixing matrix characterizing 
  the charged current interaction of Majorana neutrinos in gauge theories
  may have complex entries that conserve CP~\cite{schechter:1981hw}. These
  special CP conserving situations are associated with Wolfenstein's  
  CP-signs~\cite{Wolfenstein:1981rk}, when neutrino mass states are 
  CP eigenstates with CP parity $\eta_{\rm CP} = \pm i$. There are 
  four possible inequivalent sign configurations in the sum in 
  Eq.~(\ref{eq:meff_expl}): 
  $(+++)$, 
  $(+--)$, 
  $(+-+)$, 
  and $(++-)$. 
These are in correspondence to special values of the Majorana phases, namely
 $\phi_{12} = \phi_{13} = 0$,
 $\phi_{12} = \phi_{13} = \pi/2$, 
 $\phi_{13} = \pi/2$ with $\phi_{12} = 0$, 
and $\phi_{13} =0 $ with $\phi_{12} = \pi/2$, respectively. 
Majorana phases would also show up in processes analogous to
neutrinoless double beta decay, such as decays like $K^+ \to \pi^- \,
\mu^+ \mu^+$, whose amplitude would be proportional to $\sum U_{\mu i}^2 \,
m_i$, and have extremely low branching ratios, see Ref.~\cite{Rodejohann:2011mu} and references therein. 

Let us now comment on neutrino--anti-neutrino oscillations. 
A Gedankenexperiment looking for anti-neutrinos $\bar{\nu}_\beta$ in 
a beam of neutrinos $\nu_\alpha$ has been suggested in Ref.~\cite{schechter:1980gk} 
in order to clarify the physical nature of Majorana phases at the two-generation level.
In the three-generation case the probability for such a process is given as
\begin{eqnarray}
P(\nu_\alpha \to \bar{\nu}_\beta) & =  & 
\frac{1}{E^2} \left| \sum_j
U_{\alpha j} U_{\beta j} \, m_j \, \text{e}^{-i E_j t}
\right|^2  =  \frac{1}{E^2} 
\left| 
\sum_{i,j} U_{\alpha j} U_{\beta j}  U^*_{\alpha i} U^*_{\beta i} \,
m_i m_j \, 
\text{e}^{-i (E_j - E_i) t} \right| . 
\end{eqnarray}
leading to  complicated transition probability expressions, which will not be explicitly given here. The least 
unrealistic channel is represented by $\nu_e$ to $\bar{\nu}_e$ transitions, 
because the transition probabilities go with the ratio of mass over energy 
squared, and electron neutrinos can be produced with much lower energy than 
the other flavors.  For three families, $P(\nu_e \to \bar{\nu}_e)$ depends 
only on the Majorana phases $\phi_{12}$ and $\phi_{13}$ for the symmetrical
parametrization, whereas the PDG case leads to a dependence on all
three phases. Note the analogy to the effective mass discussed above. 

\subsection{Application to seesaw and sterile neutrinos}
  We now turn to the lepton mixing matrix characterizing models
  containing gauge singlets such as seesaw models. Their most general
  form was presented within the symmetrical parametrization in
  Ref.~\cite{schechter:1980gr}, covering seesaw schemes of all types,
  type-I, type-II, and type-III. Since it applies to an arbitrary 
  number $m$ of non-doublet leptons (singlets in type-I and II,  triplets
  in type-III) this parametrization also covers low-scale in addition to
  high-scale seesaw schemes. 
  To a good approximation, in the standard high-scale seesaw case neutrino
  oscillations are well-described by the simplest unitary form of the lepton 
  mixing matrix used above in Eqs.~(\ref{eq:pdg}) or (\ref{eq_jv}). 
  In contrast, for the low-scale seesaw
  schemes~\cite{mohapatra:1986bd,gonzalez-garcia:1989rw,akhmedov:1995ip,malinsky:2005bi}
  neutrino oscillations involve only a unitarity-violating truncation of
  the full mixing matrix, see, for example,
  Refs.~\cite{Hettmansperger:2011bt,Forero:2011pc}.
  In both cases one has an ``effective'' neutrino oscillation description 
  with $m=0$, i.e.~the extra neutral states are too heavy to take part in 
  the oscillation phenomena. 
Since these possibilities have already been widely discussed in the literature
here we will focus on the alternative possibility that singlets are
light enough to participate in oscillations, in the simple case of
$m=1$, i.e.~one ``sterile'' neutrino plus three active ones. This
possibility has recently re-gained attention~\cite{Mention:2011rk}. In
terms of the mixing matrix, a useful order of rotation is
34-24-14-23-13-12.  We have therefore
\begin{equation}
U = \omega_{34}(\theta_{34};0) \, \omega_{24}(\theta_{24},\delta_{24}) \,
\omega_{14}(\theta_{14}; \delta_{14}) \, \omega_{23}(\theta_{23};0) \,
\omega_{13}(\theta_{13}: \delta_{13}) \, \omega_{12}(\theta_{12};0) \, P  \, , 
  \end{equation}
with $P = {\rm diag}( \text{e}^{i \alpha},\text{e}^{i
\beta},\text{e}^{i \gamma})$ 
in the sense of the PDG description, or in the symmetrical form: 
\begin{equation}
K = \omega_{34}(\theta_{34};\phi_{34}) \, \omega_{24}(\theta_{24},\phi_{24}) \,
\omega_{14}(\theta_{14}; \phi_{14}) \, \omega_{23}(\theta_{23};\phi_{23}) \,
\omega_{13}(\theta_{13}; \phi_{13}) \, \omega_{12}(\theta_{12};\phi_{12}) \, . 
  \end{equation}
  First note that the number of rotations for 3 + 1 neutrino types
  (six) is exactly the number of phases (3 Dirac and 3 Majorana), a
  characteristic feature of the symmetrical
  parametrization~\cite{schechter:1980gr}.

  Consider first the effective mass characterizing \onbb in this case,
  which is given as
\begin{equation}
\meff = \left\{ \baz 
\left| c_{12}^2  c_{13}^2 c_{14}^2 \, m_1 \, \text{e}^{2i \alpha} + 
s_{12}^2  c_{13}^2 c_{14}^2 \, m_2 \, \text{e}^{2i \beta} + 
s_{13}^2 c_{14}^2 \, m_3 \, \text{e}^{2i (\gamma - \delta_{13})} 
+ s_{14}^2 \, m_4 \, \text{e}^{-2i \delta_{14}} \right|
&  \mbox{(PDG)}\, , \nonumber \\
\left| c_{12}^2  c_{13}^2 c_{14}^2 \, m_1 + 
s_{12}^2  c_{13}^2 c_{14}^2 \, m_2 \, \text{e}^{2i \phi_{12}} + 
s_{13}^2 c_{14}^2 \, m_3 \, \text{e}^{2i \phi_{13}} 
+ s_{14}^2 \, m_4 \, \text{e}^{2i \phi_{14}} \right|
&  \mbox{(symmetrical)} \, , \nonumber 
\ea 
\right. 
  \end{equation}
  Obviously the symmetrical parametrization has advantages over the
  one of PDG. Indeed, again, the CP conserving cases corresponding to
  different CP-signs are obtained by choosing the three Majorana
  phases $\phi_{12,13,14}$ to be $\pi/2$ or zero.

  In what regards oscillations, the three Dirac phases in the PDG description are
  $\delta_{13}$, $\delta_{14}$ and $\delta_{24}$. While there are nine
  \cite{Guo:2001yt} independent $J_{\alpha\beta}^{ij} = {\Im}\left\{
    U_{\alpha i}^\ast \, U_{\beta j}^\ast \, U_{ \alpha j} \, U_{\beta
      i} \right\} $, only three independent CP asymmetries
  $P(\nu_\alpha \to \nu_\beta) - P(\nu_\beta \to \nu_\alpha)$
  exist. In the symmetrical parametrization, the relevant independent
  phase combinations appearing in oscillation probabilities are \bea
  I_{123} = \phi_{12} + \phi_{23} - \phi_{13} \, , \\
  I_{124} = \phi_{12} + \phi_{24} - \phi_{14} \, , \\
  I_{134} = \phi_{13} + \phi_{34} - \phi_{14} \, .  \eea Each of the
  phase combinations $I_{ijk}$, with $i<j<k$, is seen to span three
  generations, as necessary for the existence of CP violation in
  neutrino oscillations. Note that there is a fourth possible
  combination, $I_{234}$. However, since
\begin{equation} \label{eq:sumrule}
I_{123} +  I_{134} - I_{124} = I_{234} 
  \end{equation} 
  holds, this fourth invariant is not independent. This condition
  actually implies that $I_{ijk}$ is a 2-cocycle
  \cite{Gronau:1985kx}. This is true for an arbitrary number $N_s$ of
  additional sterile neutrinos: noting that the $\frac 12 \, N_s \,
  (N_s - 1)$ rotations between sterile neutrinos are unphysical, it is
  easy to see that for $N$ massive neutrinos, including $N_s = N - 3$
  sterile neutrinos, there are $N - 1 = N_s + 2$ Majorana phases and
  $2 \, N - 5 = 2 \, N_s + 1$ Dirac phases. Each massless neutrinos
  results in one Majorana phase less. The total number of $3 \, (N -
  2) = 3 \, (N_s + 1)$ phases corresponds exactly to the number of
  physical rotations, which is $\frac 12 \, N \, (N - 1) - \frac 12 \,
  N_s \, (N_s - 1) = 3 \, (N - 2)$. The symmetrical parametrization
  is therefore tailor-made also for concisely describing the
  phenomenology of sterile neutrinos.

  It has been argued that current neutrino data might imply in fact
  that 2 sterile neutrinos are present~\cite{sorel2004combined}. A
  possible order of the nine physical rotations is 
\bea
 K =
  \omega_{25}(\theta_{25};\phi_{25}) \,
  \omega_{34}(\theta_{34};\phi_{34}) \,
  \omega_{35}(\theta_{35};\phi_{35}) \,
  \omega_{24}(\theta_{24},\phi_{24})
  \\
  \omega_{23}(\theta_{23};\phi_{23}) \, \omega_{15}(\theta_{15};
  \phi_{15}) \, \omega_{14}(\theta_{14};\phi_{14}) \,
  \omega_{13}(\theta_{13}; \phi_{13}) \,
  \omega_{12}(\theta_{12};\phi_{12}) \, .  
 \label{eq:onbb(3,2)}
\eea
Again, the effective mass characterizing \onbb automatically looks
straightforward:
\begin{equation}\label{eq:meff5}
\meff = \left|c_{12}^2c_{13}^2c_{14}^2c_{15}^2 \, m_1+
s_{12}^2c_{13}^2c_{14}^2c_{15}^2 \, \text{e}^{2i \phi_{12}}  + 
s_{13}^2c_{14}^2c_{15}^2 \, m_3 \, \text{e}^{2i \phi_{13}}
+s_{14}^2c_{15}^2 \, m_4 \, \text{e}^{2i \phi_{14}} + s_{15}^2 \,
m_5  \, \text{e}^{2i \phi_{15}}
\right|  .
\end{equation}
involving just the four physical Majorana phases.
Regarding CP violation in oscillations, there are $\left(
\ba 5 \\ 3 \ea \right) -3 = 10 - 3 = 7$ different $I_{ijk}$ combinations with
$i<j<k$ and $i,j \in \left\{1,2,3,4,5\right\}$, where the subtraction
of 3 stems from the cases which have both 4 and 5 in $ijk$: 
\begin{equation}
I_{123}, ~I_{124},~ I_{125},~ I_{134},~ I_{135},~ I_{234},~ I_{235} \, . 
  \end{equation}
There also exist $\left( \ba 5 \\ 4 \ea \right) -3 = 5 - 3 = 2$ 
``sumrules'' in analogy to Eq.~(\ref{eq:sumrule}), namely
\begin{equation}
I_{123} +  I_{134} - I_{124} = I_{234} \mbox{ and } 
I_{123} +  I_{135} - I_{125} = I_{235} \, . 
  \end{equation} 
  Hence, at the end there are 5 physical Dirac CP violating phase
  combinations affecting neutrino oscillation probabilities, for
  instance one could choose
\begin{eqnarray}
  I_{123} = \phi_{12} + \phi_{23} - \phi_{13} \, , \\
  I_{124} = \phi_{12} + \phi_{24} - \phi_{14} \, , \\
  I_{134} = \phi_{13} + \phi_{34} - \phi_{14} \, , \\
  I_{125} = \phi_{12} + \phi_{25} - \phi_{15} \, , \\
  I_{135} = \phi_{13} + \phi_{35} - \phi_{15} \, . 
 \label{eq:osc(3,2)}
\end{eqnarray}
The phase relevant for CP violation in the short-baseline oscillations
$\stackrel{(-)}{\nu_e} \leftrightarrow \stackrel{(-)}{\nu_\mu}$ sector
is ${\Im}\left\{ U_{e4}^\ast \, U_{\mu 5}^\ast \, U_{ \mu 4} \, U_{e
    5} \right\} \propto \sin(\phi_{14} - \phi_{15} - \phi_{24} +
\phi_{25}) = \sin(I_{125} - I_{124})$.

The generalization to $N_s$ sterile neutrinos is now clear: In more
mathematical language, adopted from Ref.~\cite{Gronau:1985kx}, one may
define the operator $\delta$ (with $\delta^2 = 0$) such that
(cf.~Eq.~(\ref{eq:sumrule}))
\begin{equation}
\delta I_{1234} \equiv F_{1234}^{(4)} 
= I_{123} +  I_{134} - I_{124} - I_{234} = 0 \, . 
  \end{equation}
  For general $i,j,k,l$ with $i<j<k<l$ one has $\delta I_{ijkl} =
  0$. For five active generations one has ten $I_{ijk}$ and five
  $F_{ijkl}^{(4)} = 0$, one of which can be expressed by the other
  four, for instance
\begin{equation}
F_{2345}^{(4)}  = F_{1235}^{(4)} + F_{1345}^{(4)} - F_{1234}^{(4)}  -
F_{1245}^{(4)} \mbox{ or } \delta F_{12345} = 0 \, . 
  \end{equation}
  Therefore, the standard result of $10 - (5 - 1) = 6$ Dirac phases is
  obtained.  If all generations were active, then the number of
  independent phase combinations is
\begin{equation} \label{eq:schechter}
\left(\ba N \\ 3 \ea \right) - 
\left[\left(\ba N \\ 4 \ea \right) - 
\left(\ba N \\ 5 \ea \right)
\right] = 
\left(\ba N \\ 0 \ea \right) - 
\left[\left(\ba N \\ 1 \ea \right) - 
\left(\ba N \\ 2 \ea \right)
\right] = \frac 12 \, (N - 1)  \, (N - 2) \, . 
  \end{equation} 
  One simply subtracts the number of $\frac 12 \, N_s \, (N_s - 1) =
  \frac 12 \, (N - 3) \, (N - 4)$ unphysical rotations from this
  result to obtain the already quoted $2 \, N - 5 = 2 \, N_s + 1$
  Dirac phases.  The first binomial in Eq.~(\ref{eq:schechter}) is the
  number of $ijk$ combinations with $i<j<k$, the second binomial is
  the number of sumrules between them, and the third binomial
  describes the linear relations existing between the sumrules. For
  instance, 5 active generations have $\left( \ba 5 \\ 3 \ea \right) =
  10$ different $I_{ijk}$ combinations with $i<j<k$ and $i,j \in
  \left\{1,2,3,4,5\right\}$. A number of $\left( \ba 5 \\ 4 \ea
  \right) = 5$ sumrules exist, which $\left( \ba 5 \\ 5 \ea \right) =
  1$ linear relation between them.

\section{Final Remarks}
In summary, the issue of a proper parametrization scheme for the
lepton mixing matrix will become relevant as experiments reach
sensitivity to CP violation, either of Dirac or Majorana type. 
Even more so, if present indications 
for sterile neutrinos are confirmed in upcoming experiments. 
While the form given in the PDG is just a re-writing of the
symmetrical form proposed in Ref.~\cite{schechter:1980gr}, we have
advocated here the conceptual advantages of the original symmetrical
presentation for the description of neutrino oscillations and,
especially, of lepton number violating processes like neutrinoless
double beta decay.

\begin{acknowledgments}
  This work was supported by the ERC under the Starting Grant MANITOP,
  by the Deutsche Forschungsgemeinschaft in the Transregio 27
  ``Neutrinos and beyond -- weakly interacting particles in physics,
  astrophysics and cosmology'' (WR) and by the Spanish MICINN under
  grants FPA2008-00319/FPA and MULTIDARK Consolider CSD2009-00064, by
  Prometeo/2009/091, by the EU grant UNILHC PITN-GA-2009-237920 (JV).
  We wish to thank Morimitsu Tanimoto and the organizers of the Mini
  Workshop on Neutrinos at the IPMU, where this paper was initiated.
  WR would like to thank Jisuke Kubo and the Yukawa Institute for
  Theoretical Physics at Kyoto University for hospitality during the
  YITP workshop "Summer Institute 2011", where part of this paper was
  written.
\end{acknowledgments}


\end{document}